\documentclass[aps,prl,twocolumn,showpacs,superscriptaddress]{revtex4}
\usepackage{bm, amsmath, amssymb}
\usepackage{graphicx,subfigure}
\usepackage[cp1251]{inputenc}
\usepackage[english, russian]{babel}

\begin{document}

\title{Exchange relaxation as the mechanism of ultrafast spin reorientation in
  two-sublattice ferrimagnets. }

\author{V.~G. Baryakhtar}
\affiliation{Institute of Magnetism, 03142 Kiev, Ukraine}

\author{V. I. Butrim}
\affiliation{Taurida National V.I. Vernadsky University, 95007 Simferopol, Ukraine}

\author{B.~A. Ivanov}
\email{bivanov@i.com.ua}
\affiliation{Institute of Magnetism, 03142 Kiev, Ukraine}
\affiliation{National Taras Shevchenko University of
Kiev, 03127 Kiev, Ukraine}

\begin{abstract}
In the exchange approximation, an exact solution is obtained for the
sublattice magnetizations evolution in a two-sublattice ferrimagnet.
Nonlinear regimes of spin dynamics are found that include both the
longitudinal and precessional evolution of the sublattice
magnetizations, with the account taken of the exchange relaxation.
In particular, those regimes describe the spin switching observed in
the GdFeCo alloy under the influence of a femtosecond laser pulse.
\end{abstract}

\pacs{75.10.Hk, 78.47.J-, 05.45.-a}


\maketitle

Magnetic materials have various applications in modern electronics
and informatics, but probably the most important research direction
is  still the creation of information storage and processing
systems. The  challenge of designing magnetic devices with ever
increasing information density and recording speed requires solving
certain fundamental problems of the magnetism dynamics. The
possibility to manipulate the magnetization by means of femtosecond
laser pulses opens wide opportunities in this direction. This field
has been incepted by the work \cite{[1]}, where a fast (within a
time shorter than a picosecond) reduction of nickel magnetization
after the exposure to a 100 femtosecond laser pulse has been
observed, as well as the subsequent relaxation of the magnetization
with a characteristic time of the order of picoseconds. The authors
explained the initial drop in the magnetization either by an
extremely rapid heating of the sample above the Curie point, see
review \cite{[2]}, or by spin-dependent super-diffusive electron
transfer in the laser-excited metal \cite{[12]}. Further work in
this area followed for various materials, and unexpected and rather
unusual effects were discovered. In the ferrimagnetic  rare earth
and transition metal alloy GdFeCo, a femtosecond pulse lead, in the
first stage,  to a similar spin reduction (i.e., the reduction of
the magnetization of sublattices) as for nickel, but the subsequent
evolution turned out to be fundamentally different. Instead of a
simple relaxation to the initial value, within about the same time
(a few picoseconds), both sublattice magnetizations changed their
signs, i.e., a switching of the net magnetic moment took place
\cite{[13]}, and during this picosecond-scale evolution there
occurred an \emph{a priori} energetically unfavorable state with
parallel sublattice moments. Such a magnetization switching effect
is of a threshold type, and is observed only for  sufficiently
strong pulses. It has been detected in films as well as in
microparticles \cite{[14]} and nanoparticles \cite {[15]}, both for
ferromagnets with and without a compensation point \cite{[14]}.
There is also a way of ``selective'' switching: due to the magnetic
dichroism, the absorbed energy of a circularly polarized pulse
depends on the direction of the magnetic moment of the particles,
and a pulse of certain polarization would only switch the moments of
the particles which are in a matching state \cite{[111]}. All that
makes possible to create a purely optically-controlled magnetic
memory with a picosecond recording speed.

Although an analytical explanation of this effect is highly
desirable, The theoretical description has been performed only by
means of numerical simulation \cite{[13],[14]}. It has been found
that the change of the sublattice magnetization lengths
 $S_1
=\vert {\rm {\bf S}}_1 \vert $ и $S_2 =\vert {\rm {\bf S}}_2 \vert
$, i.e. a longitudinal spin evolution, is crucially important for
this phenomenon \cite{[14],[16]}. The magnetization length is formed
by the exchange interaction, and all the salient features
(particularly, picosecond-scale characteristic evolution times, and
the fact that the effect persists even in magnetic fields up to
300~KOe) point out to the importance of the exchange-dominated
evolution \cite{[14]}.

The  Landau-Lifshitz (LL) equation \cite{[17]}, with the standard
relaxation terms \cite{[17],[18]} preserves the magnetization length.
 The problem of the correct structure of the relaxation terms
in the LL equation, including the question of a purely exchange
relaxation, was previously considered by one of the present authors
\cite{[19],[20]}. It was shown that the longitudinal spin evolution
arises naturally when the  general equations describing the
magnetization dynamics of ferromagnets \cite{[19]} and
antiferromagnets \cite{[20]} are constructed, but has certain
limitations. Because of the obvious symmetry of the exchange
interaction, it can not lead to a change (in particular, relaxation)
of the total spin of the system. Therefore, the evolution of the
magnetization length of a simple ferromagnet is reduced to a
diffusion process (that is generally nonlinear), and is absent in
the homogeneous case which we are interested in, see the detailed
analysis in \cite{[19]}-\cite{[21]}. However, for a magnet with two
sublattices, the situation is different, and a purely exchange
relaxation is possible even for a homogeneous dynamics \cite{[20]}.

These ideas were used in \cite{[16]} for a qualitative description
of the experimental data. Since the duration of the laser pulse used
(less than 100~fs) is much shorter than the characteristic evolution
time, the analysis can be performed by considering the dynamics of
the magnetization outside of the time interval of the pulse. In
doing so, a highly non-equilibrium state created by the pulse plays
the role of the initial condition for the equations describing the
magnetization dynamics. The following scheme has been proposed: the
light pulse transfers the system into a non-equilibrium state, in
which, however, the direction of the spin sublattices is the same as
in the initial state. The system evolves further under the influence
of a faster exchange relaxation, following  along the straight line
$ S_1 + S_2 = S_1 (0) + S_2 (0) = \mathrm {const} $ in the $ (S_1,
S_2) $ plane, see Fig.~1 of Ref. \cite{[16]}. The analysis showed
that the evolution of the system quickly leads to a state of partial
equilibrium, which corresponds to the spin values differing from the
initial ones $ S_1 (0)$, $S_2 (0) $ not only in the magnitude, but
in the sign as well. The further evolution is due to the slower
relativistic relaxation, and the system goes to one of the two
equivalent states of complete equilibrium. In a wide range of the
initial values, consistent with the experiment, the final state
after the two-step process differs from the initial one only by the
signs of $ S_1 $ and $ S_2 $, which explains the effect of spin
switching. However,  Ref.\cite{[16]} studied a purely longitudinal
dynamics, that is, it was assumed that the vectors $ {\rm {\bf S}}
_1 $ and $ {\rm {\bf S}} _2 $ remain collinear to their initial
values.

In the present work, the exchange evolution of  sublattice spin
vectors of a ferrite is investigated in a general way, without the
assumption of collinearity. We have found nonlinear regimes of spin
dynamics, including both longitudinal and precessional evolution of
the sublattice spins. It is shown that in the case of a strong
deviation from equilibrium an instability of the longitudinal
dynamics is possible, in which the amplitude of the precession
increases due to the transfer of the energy  associated with the
nonequilibrium character of length of the antiferromagnetism vector
$ {\rm {\bf L}}={\bf S}_{1}-{\bf S}_{2} $ into the deviation of $
{\bf L}$ from its equilibrium direction that is collinear to the
total magnetization ${\bf M} = {\bf S}_{1}+{\bf S}_{2}$.

The LL equations for a two-sublattice magnet, with
purely exchange relaxation terms can be written as
\begin{eqnarray}
\label{eq1} \nonumber \hbar \frac{\partial {\rm {\bf S}}_1
}{\,\partial t}&=&[{\rm {\bf S}}_1 {\rm {\bf ,H}}_1 ]+\lambda ({\rm
{\bf H}}_1 -{\rm {\bf H}}_2 ) -\lambda _1 \nabla ^2 {\rm {\bf H}}_1
,\\  \hbar\frac{\partial {\rm {\bf S}}_2 }{\,\partial t}&=&[{\rm
{\bf S}}_2 {\rm {\bf ,H}}_2 ]-\lambda ({\rm {\bf H}}_1 -{\rm {\bf
H}}_2 ) -\lambda _2 \nabla ^2 {\rm {\bf H}}_2 ,
\end{eqnarray}
where ${\rm {\bf S}}_1 $, ${\rm {\bf S}}_2 $ are the sublattice spins, ${\rm
{\bf H}}_{1,2} =-\delta w/\delta {\rm {\bf S}}_{1,2} $ are the effective fields
for the sublattices, and $w=w\{{\rm {\bf S}}_1 ,{\rm {\bf S}}_2 )\}$ is the
non-equilibrium thermodynamic potential per elementary cell, written as a
functional of the sublattice spin density. In what follows we  set the
Planck constant to unity, and it will only be recovered in some final results.
The relaxation terms can be written in the form $\delta Q/\delta {\rm {\bf
 H}}_{1,2} $, where $Q$ is the dissipative function, $dw/dt-=2Q$, whose
density in the exchange approximation is given by the following
expression \cite{[19]}:
\[
2Q=\lambda ({\rm {\bf H}}_1 -{\rm {\bf H}}_2 )^2+\lambda _1 (\nabla
{\rm {\bf H}}_1 )^2+\lambda _2 (\nabla {\rm {\bf H}}_2 )^2.
\]
Hereafter, we will discuss only the homogeneous dynamics, and the terms
containing  $\lambda _{1,2} $, which determine the spin diffusion, will be
neglected.

Here a general remark is in order, regarding the equations of motion
of the magnetization. For the LL equation, both dynamic and
dissipative terms (including the standard relaxation term of the
relativistic nature as well as the exchange terms such as those in
Eq.\ (\ref{eq1})) are chosen to be linear in the components of the
effective field. This approach is consistent with the Onsager
principle, see \cite{[19]}. However,  \emph {the linearity of
equations in the effective field} does not limit the applicability
of these equations to \emph{the linear approximation}. For a
magnetically ordered state, a significant nonlinearity is present in
the expression for the nonequilibrium thermodynamic potential, which
determines well-known non-linear properties of the LL equation. The
presence of this non-linearity, reflecting the properties of the
system, makes this approach very natural and reasonable. Of course,
it is possible to consider generalizations of these equations
including the terms nonlinear in the components of the effective
field, but we do not know any examples where such a generalization
would lead to new physical effects.

In the homogeneous case and in the exchange approximation, the relaxation for
two-sublattice magnets is actually determined by a single parameter $\lambda $.
This is easy to understand by noticing that Eqs.\ (\ref{eq1}) preserve the total
spin ${\bf M}$, which is the consequence of the exchange approximation. We
remark that the SU(2) exchange symmetry does not exclude the change of length as
well as the direction of the antiferromagnetism vector ${\rm {\bf L}}$.  Thus,
we come to the conclusion that the inter-sublattice exchange plays the
dominating role in relaxation (in contrast to the independent relaxation of
every sublattice, as it comes out when the relaxation term is taken in the
Gilbert form), which is supported by recent experiments \cite{[222]}.

Naturally, (\ref{eq1}) describes only the relaxation to a partially
equilibrium state, which corresponds to a minimum of the
thermodynamic potential at fixed (and, generally, non-equilibrium)
${\rm {\bf M}}$. The value of $\lambda $ can be found from the
damping decrement $\gamma _{\mathrm{lin}} $ of small-amplitude ${\rm
{\bf L}}$ oscillations, which in the framework of (\ref{eq1}) is
determined by the formula $\gamma _{\mathrm{lin}} =\lambda J_{12}
(\bar {S}_1 -\vert \bar {S}_2 \vert )^2/\bar {S}_1 \vert \bar {S}_2
\vert $, where $\bar {S}_1$, $\bar {S}_2$ are the equilibrium spin
values. It is important that the damping of optical spin waves,
connected to the transversal oscillations of the antiferromagnetism
vector ${\rm {\bf L}}$, and the relaxation of the length of ${\rm
{\bf L}}$ are both determined by the same constant $\lambda$. First,
this allows one to establish the value of $\lambda$ from independent
measurements, and second, one can use the known results of
microscopic calculations of magnon damping \cite{[26]} to estimate
it, which yields $\lambda \propto T^4$.

In what follows, our starting point will be the following expression for the
thermodynamic potential of a two-sublattice ferrite with purely exchange
symmetry, written for the homogeneous case as a function of the sublattice spins:
\begin{equation}
\label{eq2} w({\rm {\bf S}}_{1},{\rm {\bf S}}_{2})=f_1 (S_1^2 )+f_2
(S_2^2 )+J_{12} {\rm {\bf S}}_1 {\rm {\bf S}}_2 ,
\end{equation}
where $S_{1,2}^2 ={\rm {\bf S}}_{1,2}^2 $, and the exact form of the
functions $f_1 $ and  $f_2 $ is not yet specified. It is clear that
the terms containing $f_1 $, $f_2 $ do not contribute to the
dynamical part of (\ref{eq1}), and  $[{\rm {\bf S}}_{1,2} {\rm {\bf
,H}}_{1,2} ] \to \pm J_{12} [{\rm {\bf S}}_1 {\rm {\bf ,S}}_2 ]$. It
is convenient to pass to the equations for irreducible vectors ${\rm
{\bf M}}$ and  ${\rm {\bf L}}$. The equation for ${\rm {\bf M}}$
yields $\partial {\rm {\bf M}}/\partial t=0$, and for ${\rm {\bf
L}}$ one obtains the closed-form vector equation $\partial {\rm {\bf
L}}/\partial t=J_{12} [{\rm {\bf M,L}}]+2\lambda _e {\rm {\bf H}}_L
$, $\;{\rm {\bf H}}_L =-\partial w/\partial {\rm {\bf L}}$. Let us
choose the $z$ axis along the constant vector  ${\rm {\bf M}}=M{\rm
{\bf e}}_z $. In the convenient notation ${\rm {\bf L}}=L{\rm {\bf
l}}$, ${\rm {\bf l}}^2=1$ those equations take the form
\begin{eqnarray}
\label{eq4} \nonumber \frac{\partial L}{\,\partial t} &=& 2\lambda
_e ({\rm
{\bf lH}}_L ) -2\lambda _e \frac{\partial w}{\partial L}, \\
\frac{\partial {\rm {\bf l}}}{\,\partial t} &=& J_{12} [{\rm {\bf
M,l}}]+\frac{2\lambda _e }{L}[{\rm {\bf H}}_L -{\rm {\bf l}}({\rm
{\bf lH}}_L )],
\end{eqnarray}
where the dissipative term in the equation for ${\rm {\bf l}}$ resembles the
Landau-Lifshitz one. Equations for  $L$ and
${\rm {\bf l}}$, with the account taken of the specific form of the
thermodynamic potential can be also cast in the following convenient form:
\begin{eqnarray}
\label{eq5} \nonumber \frac{\partial L}{\,\partial t}&=&-2\lambda _e
(J_{12} L-\frac{\partial f_1 }{\partial S_1 }+\frac{\partial f_2
}{\partial S_2 }), \\ \frac{\partial {\rm {\bf l}}}{\,\partial
t}&=&J_{12} [{\rm {\bf M,l}}]+\frac{\lambda _e }{L}\left(
{\frac{\partial f_1 }{\partial S_1^2 }-\frac{\partial f_2 }{\partial
S_2^2 }} \right)[{\rm {\bf M}}-{\rm {\bf l}}({\rm {\bf lM}})].
\end{eqnarray}

Having written $l_x +il_y =\sin \theta \exp (i\varphi ),\;l_z =\cos \theta
$, it is easy to show that $\varphi =\omega t$, $\hbar \omega =J_{12}
M$, and at $\theta \ne 0,\pi $ vector ${\rm {\bf l}}$  precesses with a constant
frequency $\omega $, and the precession amplitude $L \sin \theta
$ changes with time because of the dissipation. It is interesting that for small
${\rm {\bf M}}$ a ``slowdown'' of this precession takes place. Thus, nonlinear
oscillations of arbitrary (not small) amplitude have the form of a precession of
${\rm {\bf L}}$ around the constant vector
${\rm {\bf M}}$, with the frequency $\hbar \omega =J_{12} M$:
\begin{multline}
\label{eq6} \nonumber {\rm {\bf L}}=L_z {\rm {\bf e}}_z +L_\bot
({\rm {\bf e}}_x \cos \omega t+{\rm {\bf e}}_y \sin \omega t), \\
L_z =L\cos \theta ,\;L_\bot= L\sin \theta ,
\end{multline}
where the quantities $L_z (t)$, $L_\bot (t)$ exhibit a dissipative evolution.
It is convenient to write down the equations in $L$, $\theta $ variables:
\begin{eqnarray}
\label{eq6} \nonumber  \frac{\partial L}{\,\partial t}&=&2\lambda
L(J_{12} -\frac{\partial f_1 }{\partial S_1 ^2}-\frac{\partial f_2
}{\partial S_2 ^2})+2\lambda (\frac{\partial f_2 }{\partial S_2
^2}-\frac{\partial f_1 }{\partial S_1 ^2})M\cos \theta , \\
\frac{\partial \theta }{\,\partial t}&=&-M\frac{2\lambda
}{L}(\frac{\partial f_2 }{\partial S_2 ^2}-\frac{\partial f_1
}{\partial S_1 ^2})\sin \theta .
\end{eqnarray}

For the sake of simplicity and physical clarity let us take $f_{1,2} $ in the
form of the Landau expansion of the form
\begin{equation}
\label{eq7} f_1 =\frac{J_1 }{4}(S_1 ^2-S_0 ^2)^2,\;f_2 =J_2
\frac{S_2 ^2}{2}.
\end{equation}

Here we assume that the second sublattice consists of paramagnetic
rare-earth ions, $f_2 $ is determined by the spin entropy, and $J_2
$ is of the order of the temperature $T$. The parameter $S_0 =S_0
(T)$ formally coincides with the equilibrium value of the iron
sublattice magnetization if one neglects
 its interaction with the rare-earth sublattice. Using
(\ref{eq6}), one obtains simple closed formulae for the equilibrium values of
 the sublattice spins, $\bar {S}_1
=\sqrt {S_{1,0} ^2+J_{12} ^2/J_1 J_2 }$ and $ \bar {S}_2 =-J_{12} \bar
{S}_1 /J_2 $, while the equations can be written in the form
\begin{equation}
\label{eq8} t_0 \frac{\partial L}{\partial t}=f(L,\theta ),\;t_0
\frac{\partial \theta }{\,\partial t}=g(L,\theta
),\;t_0=\frac{4\hbar}{\lambda J_1},
\end{equation}
где $ f(L,\theta )=-L^3-3L^2M\cos \theta +AL+B$,
\begin{multline}
\label{eq7} \nonumber  g(L,\theta )=-\frac{M\sin \theta}{L}
(\frac{4J_2 }{J_1 }+4S_{0,1}^2 -L^2-2LM\cos \theta -M^2) \\
 A=-M^2(1+2\cos ^2\theta )-\frac{4J_2 }{J_1 }+4S_{0,1}^2
+\frac{8J_{12} }{J_1 }, \\   B=M\cos \theta (\frac{4J_2 }{J_1
}+4S_{0,1}^2 -M^2).
\end{multline}

It is worth noting that the evolution of  $L_z (t)$, $L_\bot (t)$
occurs on a naturally emerging universal time scale $t_0 =4\hbar
/\lambda J_1 $, which is larger than the ``purely exchange'' time
$t_{\mathrm{ex}} \sim \hbar /J_1 \sim t_0 /\lambda $ since the
relaxation constant $\lambda $ is small. For not very small values
of $M\sim 1$ and not too weak inter-sublattice interaction $J_{12}
\sim J_1 $, this time scale is also larger than the precession
period of vector ${\rm {\bf L}}$.

Proceed further to the analysis of the evolution of $L_z $ and $L_\bot $. It is
clear that all singular points occur at $\theta =0,\pi $, and their positions
are determined by zeros of the function $f(L,\theta )$ at $\sin \theta =0$.
The condition $f(L,\sin \theta =0)=0$ can be represented as a cubic equation in
$L\cos \theta =\pm L$ (it is convenient to assume that $L>0$, and $\theta $
varies in the range $0\le \theta \le \pi $).
At $M=0$ the three roots are  $L_{1,3} \cos \theta =\pm \sqrt A $ and
$L_2 =0$, so it is clear that at sufficiently small  $M\le M_c $ there will also
be three real roots. A simple analysis shows that  $L=L_1 ,\;\theta
=0$ (or $L_z =L_1
>0$, $L_1 =\sqrt A $ at $M=0)$ corresponds to the equilibrium position
(a stable node), $L=L_3 ,\;\theta =\pi $, i.e.,  $L_z =-L_3 <0$ corresponds to a
saddle point, and the unstable node lies at $L=L_2 $. For
$M\le M_c $ one has $L_1 <L_3 $, and the unstable node will correspond to a
negative value of $L_z = L_2 \cos \theta <0$. At $M=M_c $ the $L_2 $ and  $L_3 $
roots merge, and for $M>M_c $ the system has only one singular point at $L\cos \theta =L_1 >0$.

\begin{figure}[tb]
\includegraphics[width=0.47 \textwidth]{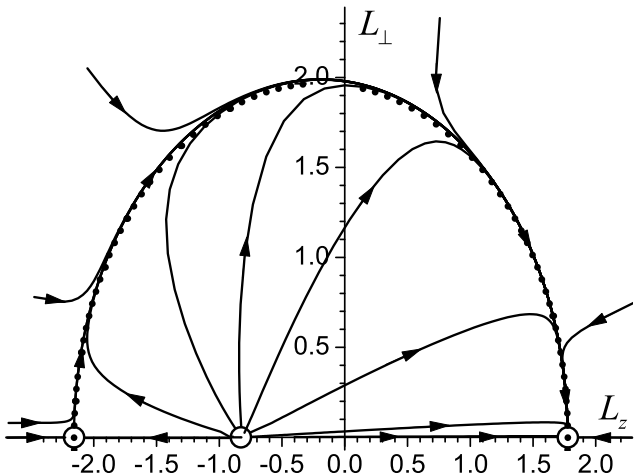}
\caption{\label{fig:1} The evolution of $L_\bot$ and $L_z$, calculated numerically for $J_1 =J_2 =2J_{12} $ and $S_0 =1$ at
$M=0.4$ and shown as a phase portrait. Singular points are shown with circles,
  and the separatrix is shown as a dotted line.  }
\end{figure}

It is important to note that for all values of  $M$ the system (\ref{eq8}) has
another solution $L_z =L_z (t)$, $L_\bot (t)=0$, which corresponds to a purely
longitudinal evolution, but the physical sense of this solution is very
different. At  $M>M_c $, for all initial conditions  ($L(0)<L_1 $ or
$L(0)>L_1 $),  the value of $L_z $ tends to its equilibrium value $L_1 $ for
this class of solutions. Numerical analysis shows that in his case the evolution
remains close to the purely longitudinal one even if the direction of ${\rm {\bf
    L}}$ deviates from the equilibrium. The only exception is for large
deviations, when $L(0) \sim -L_1 $; in this case the length of
 ${\rm {\bf L}}$ is already close to the equilibrium, and a rotation of ${\rm
  {\bf L}}$ becomes favorable. At extremely small $M$ the evolution is
degenerate: $\theta $ changes much slower than $L$, and the phase portrait in
the $(L_\bot ,L_z)$ plane consists of radial straight line intervals  $\theta
=\mathrm{const}$ and of parts of the circle $L=L_1 \approx \sqrt A $. For finite
$M<M_c $ the situation is much more interesting: in this case one also has a
solution  of the form  $L_z =L_z
(t),\;L_\bot (t)=0$, but with the initial conditions  $-L_3 <L_z (0)<-L_2
$, i.e., between the saddle point and the unstable node, the longitudinal
evolution takes the system away from the equilibrium. This is illustrated in
Fig.\ \ref{fig:1}, which shows the phase portrait of the system in $(L_\bot
,L_z)$ plane, calculated numerically for $J_1 =J_2 =2J_{12} $ and $S_0 =1$ at
$M=0.4<M_c $ (at those parameter values one has $M_c =4/(3\sqrt 3) \simeq
0.77$).

Thus, the exact solution of the full system of equations of motion for
sublattice spins  in the exchange approximation shows the existence of two
qualitatively different regimes. The character of the evolution is mainly
determined by the initial value of the magnetization
$M$, which is conserved in the exchange  approximation.
For large $M>M_c $, as well as for all $M$ and the initial value $L_z
(0)>0$, a longitudinal relaxation occurs, as studied previously in
\cite{[16]}. For $M<M_c $, approximately the same behavior is also retained for
negative $L_z (0)$, provided that $L_z (0)>-L_2 $.
In all those cases, there is a special solution of the form
 $L_z = L_z (t)$ which leads to the equilibrium, and even for a nonzero (but
small) value of the transversal initial deviation $L_\bot (0)$ the
value of $L_\bot $ remains small in the process of relaxation.
However, the situation is changed dramatically, if the initial value
enters the region of the unstable node situated around $L_z \simeq
-L_2 $ ($L_z \simeq 0.83$ in Fig.\ \ref{fig:1}). As seen in Fig.\
\ref{fig:1}, in the vicinity of this point and to its left, even
small initial values of $L_\bot $ increase with time. In this case,
in a wide range of the initial conditions all trajectories in the
$(L_z,L_\bot) $ plane tend to the separatrix which connects the
saddle point and the unstable node; the values of $L_\bot $ are not
small at the separatrix. In this way, strongly nonlinear evolution
regimes with $L_\bot \sim L_z \sim 1$ become possible. The solution
with  $L_\bot \ne 0$ at $M\ne 0$ is of the precession type, i.e.,
for the initial condition with
 $L_z <-L_2 $
approaching equilibrium is accompanied by the growth of the precession amplitude
of ${\rm {\bf L}}$, at the constant precession frequency $\hbar
\omega =J_{12} M$, so that
 ${\rm {\bf L}}=L_z {\rm {\bf e}}_z
+L_\bot ({\rm {\bf e}}_x \cos \omega t+{\rm {\bf e}}_y \sin \omega
t)$.
It should be remarked that the experimentally observed time dependence of the
sublattice magnetizations  in the time interval between $0.5$ and $3$~ps shows
some non-monotonic behavior at the background of a smooth magnetization change,
which resembles oscillations with the period of about $0.3$~ps, see Fig.~2 of
Ref.~\cite{[13]}. The results of numerical simulation of this process, reported
in the same work \cite{[13]}, did not show such a behavior, but oscillations
were found in recent numerical studies \cite{[888]}.

Taking into account the transversal spin deviations in the process
of evolution may be important for understanding the recent
experiment on TbFeCo alloy \cite{[TbFe]}. An obvious difference
between this material and GdFeCo is the presence of a strong
easy-axis anisotropy, but it is clear that such anisotropy should
not affect a purely longitudinal evolution. Despite that, spin
switching characteristic for  GdFeCo was not observed in  TbFeCo,
although the initial reduction of the sublattice magnetizations was
roughly the same as in the GdFeCo experiment. Of course, there could
be other reasons for such a different behavior, e.g., the presence
of unquenched orbital moment of Tb, but the detailed analysis of
this problem is beyond the scope of the present paper.

This work is partly supported by the joint Grant 0113U001823 of the
Russian Foundation for Fundamental Research and the Presidium of the
National Academy of Science of Ukraine, and by  the joint Grant
$\Phi$53.2/045 of the Russian and Ukrainian Foundations for
Fundamental Research.

\end{document}